# Setting Boundaries for Statistical Mechanics

Bob Eisenberg

December 21, 2021

Department of Applied Mathematics
Illinois Institute of Technology;

Department of Physiology and Biophysics
Rush University Medical Center
Chicago IL
*Bob.Eisenberg@gmail.com*

# Abstract

Statistical mechanics has grown without bounds in space. Statistical mechanics of point particles in an unbounded perfect gas is commonly accepted as a foundation for understanding many systems, including liquids like the concentrated salt solutions of life and electrochemical technology, from batteries to nanodevices. Liquids, however, are not gases. Liquids are filled with interacting molecules and so the model of a perfect gas is imperfect. Here we show that statistical mechanics without bounds (in space) is impossible as well as imperfect, if the molecules interact as charged particles, as nearly all atoms do. The behavior of charged particles is not defined until boundary structures and values are defined because charges are governed by the Maxwell partial differential equations. Partial differential equations require boundary conditions to be computable or well defined.

The Maxwell equations require boundary conditions on finite sized spatial boundaries (i.e., structures). Boundary conditions cannot be defined 'at infinity' in a general (i.e., unique) way because the limiting process that defines infinity includes such a wide variety of behavior, from light waves that never decay, to fields from dipole and multipolar charges that decay steeply, to Coulomb fields that decay but not so steeply. Statistical mechanics involving charges thus involves spatial boundaries and boundary conditions of finite size. Nearly all matter involves charges, thus nearly all statistical mechanics requires structures and boundary conditions on those structures.

Boundaries and boundary conditions are not prominent in classical statistical mechanics. Including boundaries is a challenge to mathematicians. Statistical mechanics must describe bounded systems if it is to provide a proper foundation for studying matter.

# Introduction

Molecular systems nearly always involve electrical properties, because matter is held together by electrical forces, as specified by quantum chemistry. The role of electrical forces is obvious from the Schrödinger wave equation of the electron, that specifies quantum chemistry. The Schrödinger equation includes the electrical potential $V$. The Bohm formulation of quantum mechanics illustrates the role of electrodynamics in familiar ways.[2-4] The Hellmann Feynman theorem makes this role of electricity more explicit as the source of forces in atoms and molecules.[5, 6]

Even uncharged atoms like argon interact through the quantum fluctuations of their charge density, which are stochastic deviations from the mean charge density of zero. These London dispersion forces are electrical. They are important determinants of macroscopic forces.[7]

The unavoidable conclusion then is that theories, calculations, or simulations of molecules must satisfy the laws of electrodynamics.

The question then is what are the laws of electrodynamics that molecular simulations and statistical mechanics must satisfy?

**<u>Why isn't electrostatics good enough?</u>** Molecular and atomic simulations use Coulomb's law to describe electrical forces.

Coulomb's law is a simple algebraic law that does not include time. It is a static description of electrodynamics and as such obviously cannot describe the dynamics of charges and fields.

$$\text{Electrical Force} = \frac{1}{4\pi\varepsilon_0} \frac{q_1 q_2}{r^2} \tag{1}$$

where the charges $q_1$ and $q_2$ at distance $r$ produce the electrical force with electrical constant $\varepsilon_0$, the permittivity of free space.

Sadly, electrostatics cannot provide a sound foundation for statistical mechanics because charge moves rapidly on the atomic scale of Ångstroms and femtoseconds. Electrostatics obviously is not enough to describe femtosecond events. Electrodynamics is needed.

Fig. 1.

**Feynman's language is imperative:**

1) "**Coulomb's law … is to be used only for statics**."

2) "**What is True for Statics is False for Dynamics**"
   Title and Contents from Feynman's Section 15-6  [1]

Feynman's Table 15-1 presents eleven equations that are labelled
3) "**FALSE IN GENERAL (true only in statics)**"
   Upper case in original.
   **Coulomb's law**
   is labelled **false**.

**Feynman's impatience with the misuse of Coulomb's law**
seems obvious from his repeated choice of imperative language.

Fig. 1. The need for electrodynamics, not just electrostatics, is emphasized by Feynman [1], in language that could hardly be more explicit.

# Theory

**The laws of electrodynamics are the Maxwell equations.** The Maxwell equations (as written by Heaviside and others) are universal laws valid over an enormous range of times and distances.

## Core Maxwell Equations

$$\text{div } \mathbf{E} = \frac{\rho}{\varepsilon_0} \tag{2}$$

$$\text{div } \mathbf{B} = 0 \tag{3}$$

$$\text{curl } \mathbf{E} = -\frac{\partial \mathbf{B}}{\partial t} \tag{4}$$

$$\text{curl } \mathbf{B} = \mu_0 \left( \mathbf{J} + \varepsilon_0 \frac{\partial \mathbf{E}}{\partial t} \right) \tag{5}$$

The core Maxwell equations (1) – (4) use the variables $\rho$ and $\mathbf{J}$ to describe all charges, however small, and all flux (of charges with mass), however fast, brief, and transient. Equations (1) – (4) are called the core Maxwell equations because they include polarization phenomena in the properties of $\rho$ and $\mathbf{J}$ rather than in the conventional way, shown in Fig. 2.

Nonlinear terms are not present in the Maxwell equations. Nonlinearity has not been observed in experiments, but is predicted at very large field strengths, approaching the Schwinger limit of some $1.32 \times 10^6$ volts/Ångstrom.

Fig. 2

**Cartoon of Reassignment of Polarization as part of J and $\rho$**

*Core Maxwell*            *Conventional Maxwell*

$$\mathbf{div\ E} = \frac{\rho}{\varepsilon_0} \qquad\qquad \mathbf{div\ }D = \mathbf{div\ }\varepsilon_r\varepsilon_0\mathbf{E} = \rho_f \tag{6}$$

$$\frac{1}{\mu_0}\mathbf{curl\ B} = \mathbf{J} + \varepsilon_0 \frac{\partial \mathbf{E}}{\partial t} \qquad\qquad \frac{1}{\mu_0}\mathbf{curl\ B} = \tilde{\mathbf{J}} + \varepsilon_r\varepsilon_0 \frac{\partial \mathbf{E}}{\partial t} \tag{7}$$

Fig. 1. Classical and Core Maxwell equations. $\tilde{\mathbf{J}}$ describes the flux of mass with charge, after the usual dielectric term is subtracted from $\mathbf{J}$. $\rho_f$ describes the distribution of charge after the usual dielectric term is subtracted from $\rho$.

**Classical Maxwell equations are constitutive equations.** They embed the dielectric constant of matter into the very definition of their variables. For example, the Maxwell vector field **D** is defined to include the polarization **P** of matter and the relative dielectric constant $\varepsilon_r$, a single dimensionless positive real number, sometimes called the relative permittivity. $\varepsilon$ is the (dimensional) permittivity. The electric susceptibility is $\chi = \varepsilon_r - 1$.

$$\mathbf{D} \triangleq \varepsilon_0 \mathbf{E} + \mathbf{P} = \varepsilon_0 (\varepsilon_r - 1) \mathbf{E} \tag{8}$$

$$\mathbf{D} \triangleq \varepsilon_0 \mathbf{E} + \mathbf{P} = \chi \mathbf{E} = \varepsilon \mathbf{E} \tag{9}$$

The classical Maxwell equations therefore must be revised, into quite different form in fact, when the dielectric constant is not constant, that is to say, when the polarization cannot be described by a single real positive number $\varepsilon_r \geq 1$.

In fact, the description of polarization by a single positive real number is almost never an adequate representation of the properties of real systems.[8-11] The reformulation of the Maxwell equations for nonconstant $\varepsilon_r - 1$ will produce equations with very different mathematical form, in general requiring convolutions in the time domain, although with similar physical content.

Of course, when nothing is known experimentally about polarization **P**, it is better to use a dielectric description with $\varepsilon_r$ constant, than with no polarization $\mathbf{P} \cong \mathbf{0}$, at all.

**Maxwell core equations are not constitutive equations.** The core equations contain no parameters describing matter. The core Maxwell equations involve only two parameters, and those are parameters of space, not matter: the magnetic parameter (i.e., permeability of space) $\mu_0$, and electric parameter (permittivity of space) $\varepsilon_0$, and perhaps the speed of light $c = 1/(\mu_0 \varepsilon_0)^{1/2}$. These parameters are true constants within the accuracy (~$10^{-8}$) of measurements of the fine structure constant α of quantum electrodynamics. They are universal field equations true everywhere, in the vacuum of space and in matter, including in the vacuum within and between atoms.

The core Maxwell equations may seem to be quite useless without a specific description of material charge. Indeed, they are useless if the goal is a complete description of electrodynamics.

If the goal is to describe the properties of (total) current, however, the core Maxwell equations are remarkably useful, even without knowledge of constitutive properties.

**Kirchhoff's Current Law does not depend on Constitutive Properties.** The core equations allow the derivation of a form of Kirchhoff's current law that is as precise as the Maxwell equations themselves.[12, 13] And Kirchhoff's current law is the main (and often the only) theory needed to define and design the electronic circuits of our digital technology, as reference to texts of circuit design shows most eloquently.[14-20] Charges are hardly ever discussed in circuit design, usually only in the charge of parasitic capacitance [13], which turns out to include and sometimes be the same as the charge described by the ethereal current $\varepsilon_0\, \partial \mathbf{E}/\partial t$. I use the name 'ethereal current' to describe Maxwell's displacement current in deference to the enormous role that the aether (and its current $\varepsilon_0\, \partial \mathbf{E}/\partial t$) had in the history of electrodynamics [21-30] despite its relegation to its present day ghostly nonphysical role by the theory of relativity.[26, 31-34]

When matter is involved, and a complete description is the goal, a separate constitutive theory of charged matter—'material charge' I like to call it—is needed, as we shall see, to describe how charge $\rho$ and the flux **J** change as the electric field changes.

**Constitutive theory of charged matter is rather similar to the constitutive theory of mass**. Polarization can be described by constitutive equations. The stress strain relations of solids are a constitutive theory of mass, as are the stress strain relations of complex fluids.[35-38]

**Polarization**. The distribution and amount of charge in matter varies with the electric field. Charge is said to polarize in the electric field. Solid matter polarizes. So do liquids, as seen in the literature under the name 'concentration polarization'.[39-42] Hodgkin, Huxley, and Katz [43] leave out the modifier 'concentration', and say 'polarization' thus making confusion with dielectric polarization **P** inevitable, particularly for biologists not as familiar with the history of electrodynamics as they might wish to be.[21, 22, 26, 29] Molecules polarize in complex, time and field dependent ways. So do atoms, and of course aggregates of molecules, as reported in the literature of impedance, dielectric and molecular spectroscopy.[44-55] Polarization of proteins is one kind of the conformation changes of proteins used so widely to describe their function. Conformation changes occur over an enormous range of times scales in proteins. So does polarization.

**Forces Change Distributions of Mass and Charge.** It is obvious that a mechanical force applied to a mechanical system changes the distribution of mass. It should be obvious that electrical force applied to a system of charges changes the distribution of charge.

As the electric field changes, forces change the amount and location of charge, much as a mechanical forces (stress) change (strain) the amount and location of mass. A description of the change of distribution of mass is likely to be quite specific to the system being studied. The description will depend on the structure within which the matter (and thus the field equations) are embedded, and on the boundary conditions that describe the location and properties of the boundaries and structures. Generalities are likely to be too vague to be very useful in applications because applications almost always depend on the shape of the structure containing the force fields and its boundary conditions.

Similarly, it should be obvious that an electrical force applied to a charged system changes the distribution of charge. And a description of the change of charge distribution is likely to be quite specific to the system being studied for the same reasons.

**Complex fluids.** The stress strain formalism of complex fluids is a powerful and general way of describing stress strain distributions of mass. In its variational form [35-38], the stress strain formalism accommodates diffusion and convection that are so important in liquids. The variational form allows the large movements of convection and diffusion in liquids to be described, along with the much smaller movements of elasticity of solids.

Similarly, the stress strain formalism of polarization can accommodate the diffusion, migration, and convection of charge in solutions in much the same way

# Results

**Polarization can be treated as the stress strain relation of charge** (see eq. 3.1–3.5 of ref. [56]). In its variational form, the stress strain formalism accommodates diffusion and convection that are so important in liquids, yielding the classical Poisson Nernst Planck equations in special cases [35-38] important in applications ranging from ions in water solutions, ions in protein channels, to ions in gases [57] and plasmas [58-60], to holes and electrons that are the quasi-ions of the semiconductors of our computers and smartphones.[61-65]

It is obvious that stress strain relations are hard to summarize. They usually involve a multitude of parameters chosen to describe the specific properties that determine the deformation of matter.

It should be obvious that the stress strain relations of charge will be at least as hard to summarize as those of mass. Those polarization properties will involve a multitude of parameters chosen to describe the deformation of distribution of charge by electric forces. A single dielectric constant will hardly ever be adequate, despite its historical provenance.[55, 66] Of course, when nothing is known experimentally about polarization, it is better to use a dielectric description than nothing at all.

Once polarization is separated from the core Maxwell equations, it is clear that the core equations are fundamental, universal and as exact as any in science.[8] Without polarization, the Maxwell core equations have only two constants and these are ***not*** adjustable. These constants are known to be just that. They are constant and do not change in any known experimental conditions. They are determined with great precision by any two of the experimentally determined properties, the electrical constant $\varepsilon_0$ (the permittivity of free space), the magnetic constant $\mu_0$ (the permeability of free space), and speed of light $c$.

**Maxwell equations require boundary conditions on a finite structure**. Maxwell equations of electrodynamics are partial differential equations that require boundary conditions specified on a finite–***not infinite***–structure, called a domain in mathematics, as we shall discuss at length later in this paper, on p.11. After the boundary and conditions are specified, the size of the boundary structure, and the domain it contains, can be increased 'to infinity' to see if a unique boundary condition at infinity is possible, independent of shape, location and parameters.

It is clear that electrodynamic phenomena 'at infinity' are so diverse that they cannot be specified in a general way. The phenomena of electrodynamics include light that propagates from the edge of the universe over billions of years and the decaying phenomena of electrostatics determined by (for example) Coulomb's law. The Maxwell equations must be solved in a finite structure if the solutions are to cover the range of electrodynamic phenomena involved on the time scales of atomic motion.

Statistical mechanics and thermodynamics of matter must include electrodynamics because charges are everywhere in matter. As we have seen, interactions of even (nominally) uncharged atoms like argon involve transient charges.

Thus, **statistical mechanics and thermodynamics must be specified in finite domains. That is the main point of this paper.**

We turn now to a more detailed presentation of these same issues.

**Maxwell Equations are true on all scales**. The Maxwell equations have properties that are not common in scientific theories, and these need to be understood explicitly as we seek firm foundations for our theories and simulations.

For example, the Maxwell equations in general, and the Poisson version of Gauss's law (Maxwell's first equation eq. (2)) are often treated as averaged or mean field theory equations in my experience, perhaps because of the enormous variations of potential (say 1 electron−volt or 40 times the thermal energy) in a few picoseconds. Faced with this much variation, scientists are likely to think that equations describing potential must be averages. That is not true. The Maxwell equations describe potential as it varies during thermal motion. They describe potential as a function of time on the atomic time scale of $10^{-15}$ seconds and much faster, even much faster than the electron time scale of say $10^{-19}$ sec of quantum chemistry. The core Maxwell equations are not mean field theories or averaged in any sense.

These properties are apparent when the Maxwell equations are applied to a vacuum where $\mathbf{J} = 0$ and $\rho = 0$. Indeed, this application was historically central in Maxwell's theory of electricity and the equations that describe it.[25] In the vacuum, the source of the magnetic field $\mathbf{B}$ is the ethereal displacement current $\varepsilon_0 \, \partial \mathbf{E}/\partial t$ (because $\mathbf{div\,B} = 0$). Currents and perhaps charges found on structures that bound the vacuum region can also be sources of the magnetic field.

The ethereal displacement current $\varepsilon_0 \, \partial \mathbf{E}/\partial t$ is universally present in matter and in a vacuum, because it arises from the relativistic invariance of charge with velocity, as described in textbooks of special relativity [67], in Einstein's original paper [68, 69], or memorably in Section 13-6 of Feynman's textbook [1].

**Ethereal Currents.** The implications of the ethereal term $\varepsilon_0 \, \partial \mathbf{E}/\partial t$ are profound. The Maxwell equations involve (total) current flow and $\mathbf{E}$ fields in all of space, and cannot be confined to atoms in atomic resolution simulations. The Maxwell equations describe electric fields in discrete simulations of atoms because $\varepsilon_0 \, \partial \mathbf{E}/\partial t$ exists everywhere in those simulations, as it does everywhere in space, even if all charges are confined to atoms. Derivations of statistical mechanics must include the same realities as simulations of atoms and so are subject to the same issues.

The Maxwell equations are not confined to continuum descriptions of charge. If the 'outside the atoms' currents and fields in space are ignored, the simulations cannot satisfy the equations (2)–(5) everywhere and at every time, as physicists say they should whenever electricity is involved. Derivations of statistical mechanics must include the same realities of electrodynamic fields as simulations of atoms and so are subject to the same issues.

Mean field or low resolution models of charge may indeed be averaged. In those cases, where $\mathbf{J} \neq 0$ and $\rho \neq 0$, averaging is present. But the averaging occurs within the models of $\mathbf{J}$ and $\rho$, not in the Maxwell equations themselves. For example, averaging is usually found in the theories and simulations of polarization, e.g., it occurs in the stress strain theories of the distribution of charge and matter we have discussed.[56]

To summarize this section: $\varepsilon_0\, \partial \mathbf{E}/\partial t$ cannot be avoided even in atomic simulations. This fact often surprises colleagues used to thinking of electricity as the properties of charged atoms, and their movement.

But electricity is much more than charges and their movement. It includes all the properties of light and electromagnetic radiation everywhere.

**Electricity always includes the ethereal displacement current term** $\varepsilon_0\, \partial \mathbf{E}/\partial t$. Without $\varepsilon_0\, \partial \mathbf{E}/\partial t$, there is no source for **curl B** in a vacuum or in the space between atoms devoid of mass between or within atoms (assuming no currents on boundary structures), and light cannot exist or propagate.

The ethereal term does not depend on the properties of matter. It in fact is a property of space, not matter, arising from the fact that charge is Lorentz invariant in any locally inertial reference frame as discussed in textbooks of special relativity, in Einstein's original paper [68, 69], or memorably in Section 13-6 of Feynman's textbook [1]. Charge (unlike length, time, and relativistic mass) does not change as charges move, no matter how fast they move, even if they move at speeds close to the speed of light (as in the synchrotrons of a say 7 gigavolt advanced photon source).

**Electrodynamics requires differential equations**. The existence of the ethereal current $\varepsilon_0\, \partial \mathbf{E}/\partial t$ means that any description of electrodynamics must include a partial derivative with respect to time, usually in the form of the Maxwell equations (2) – (5). These are partial differential equations and so they cannot be computed, even approximately, without boundary conditions on their limiting structures, and initial conditions. In the language of mathematics, the solutions to the equations do not exist without boundary and initial conditions.

**Electrodynamics and the Maxwell Equations are relevant to biology.** It is natural for biologists and biochemists to think that the previous discussion is irrelevant to their concerns. One might think that ethereal displacement currents are small and so unimportant. But that is not the case as the simplest estimates show, and as can be measured in every simulation of molecular dynamics. Those always include atomic time scales in which the ethereal current is large.

Indeed, Langevin and Brownian models are often used as supplements to all-atom molecular dynamics. Those coarse grained Langevin and Brownian models include a noise term that is a Brownian stochastic process in the language of probability theory and have infinite variation, in the language of mathematics, which means that they have infinite velocity. Their noise trajectories cross a region an infinite number of times in any time interval however brief.[70-72] While it is not clear how to compute the ethereal current of charges moving this way at infinite velocity [72], it is clear that the ethereal current of a process with infinite velocity cannot be small. Indeed, it is quite likely to be large, beyond easy comprehension.

The idea of an ethereal current should not be strange. Tthe concept of ethereal current arises naturally in high school physics. It is implicit in most elementary discussions of capacitors in which the charge $Q_{cap} = C_{cap}V$ and current $I_{cap} = C_{cap}\, \partial V/\partial t$. The idealized capacitors most of us studied in elementary physics classes have large current flows in the empty space between the plates of the capacitor and that current is the ethereal current. No material charge exists or flows there. The $\varepsilon_0\, \partial \mathbf{E}/\partial t$ term is in fact the only current between the plates of a vacuum capacitor. The ethereal current is always exactly equal to the total current flow in the wires on either side of the capacitor, because total current is conserved exactly by Maxwell's equations.[73]

A vacuum capacitor may seem an artificial schoolchild example, although not to those of us who have wired up circuits with capacitors or to the thousands of circuit designers who include them in the many billions of circuits in our computers. And systems certainly exist for which $\varepsilon_0\, \partial \mathbf{E}/\partial t$ is unimportant, e.g., in many systems in which $\partial \mathbf{E}/\partial t = 0$.

But the ethereal current is almost never small for atomic scale systems, even at temperatures near absolute zero. It cannot be safely ignored in simulations or derivations of statistical mechanics that involve the atomic scale.

**Defining Infinity**. Another issue seems abstruse mathematics, but is not. Defining infinity is not quite the arcane point of pure mathematics it might seem to be. In fact, the idea of 'boundary conditions at infinity' is useful only if that phrase defines a wide class of structures far away from the system of interest, ***in which the details of the structures are unimportant*** because they are lost in the blur and haze of distance, as the details of the structure are lost in the word 'infinity'.

If different structures produce different boundary conditions when the structures are far away, a single word and equation 'at infinity' will not be able to describe the resulting range of behaviors. In fact, 'infinity' cannot be defined in a unique way from the Maxwell equations themselves as the following example shows.

Consider two subsets of the Maxwell equations. Consider an electrostatic problem, with all charge in a finite region. Coulomb's law eq. (1) can then be used to compute electrical forces. Magnetic forces do not exist (because it is a static problem). Infinity can be defined easily and uniquely and the potential or electric field at infinity is zero in electrostatic systems with charges all in one region. (If charges are not confined to one region, difficulties of convergence occur.)

But consider a different structure described by Maxwell equations in which wave properties predominate in a pure vacuum without matter. Two relevant wave equations in this domain are derived in textbooks of electrodynamics and discussed in [74].

$$\mu_0 \varepsilon_0 \frac{\partial^2 \mathbf{E}}{\partial t^2} - \nabla^2 \mathbf{E} = 0 \tag{10}$$

and

$$\mu_0 \varepsilon_0 \frac{\partial^2 \mathbf{B}}{\partial t^2} - \nabla^2 \mathbf{B} = 0 \tag{11}$$

The solutions to these equations do not go to zero at infinity. In fact, these solutions never remain close to zero. The solutions describe light waves that propagate forever, as light actually does propagate over billions of light years of distance, from galaxies at the edge of the observable universe for very long times. The light from the galaxy **GN-z11** started soon after the universe began some $1.3 \times 10^{10}$ years ago billions of light years from the earth where we observe it.

## Discussion

The analysis of 'at infinity' shows in a mathematically precise way that the Maxwell equations do not have a single set of boundary conditions 'at infinity'. Rather, each application of the Maxwell equations requires an explicit definition of confining (as well as internal) structures and the boundary conditions on those structures. It also requires a statement of how structures and conditions vary as the system gets bigger and bigger, to infinity. One description of structures and boundary conditions cannot be enough, no more at infinity than anywhere else.

Thus, any description of electrodynamic phenomena in systems get large without limit needs to specify

(1) the structure of the system
(2) the boundary conditions on the confining structure that bounds the system
(3) the change in shape of the structure as it moves 'to infinity'
(4) the change in boundary conditions as the structure moves 'to infinity'

**Statistical Mechanics unbounded**. What are we then to make of the fact that most treatments of statistical mechanics do not include boundary conditions?

Surely the results of these analyses must have value even if they are unable to include the Maxwell equations!

Of course, classical statistical mechanics has immense value. In my view, the classical results serve as a first model, from which to construct other more refined models. The more refined models can include structures and boundaries that are allowed to move to infinity.

In this view, classical statistical mechanics provides an admirable starting point for the iterative social process we call science. Statistical mechanics provides a first iterate for the handling of statistical properties of idealized, albeit impossible, systems. Later iterations provide the improvements that allow charge and the equations that describe charge. Those equations include the structures that bound the charge and the conditions on the equations at those structures.

But we must allow the scientific process to iterate if it is to improve. We must extend statistical mechanics to include structures and boundary conditions. We must remember that statistical mechanics without spatial bounds has logical bounds. It is not a universal set of laws. Statistical mechanics is a model that must like all other scientific models be compared to experiments. Those experiments include structures and bounds.

What is clear is that boundaries must be included in the final iterates of our theories and simulations of the statistical mechanics of matter, because matter is charged. Matter is charged by the Maxwell and Schrödinger equations and they are bound to include boundary conditions. They are confined by structures that form spatial boundary conditions as are all partial differential equations.

**Statistical Mechanics within Boundaries**. The inclusion of structures and boundary conditions in statistical mechanics is likely to require extensive investigation of specific problems [75] and these will not be easy to study, judging from work in related fields for example, the theory of granular flow [76-78] and soft matter [79, 80]. Specifics are needed because specific problems involve specific structures and the structures can be as important as the field equations themselves. It is the structure of the "piston in a cylinder" that converts the combustion of gasoline into motion. The field theory of combustion is silent about the motion without the structure. Each structure will then need separate investigation and general theories will tend to be less useful than one would wish.

A simple example shows the profound nature of the issues involved. Consider triangular objects ('molecules' in a flatland) in a two dimensional universe in a triangular domain.[81] It is obvious that if the triangles are similar, i.e., have the same shape, the triangles can lock, they can jam into an immobile array nearly crystalline in nature. This jamming can occur no matter how large the system, no matter how far away is the boundary. These issues are well recognized in the specialist literature of granular flow [76-78] but their remedy is unclear, not yet at hand.[80, 82, 83] It seems necessary to consider such situations as one tries to design a statistical mechanics that respects the finite bounds of real systems.

Meanwhile, one can proceed in an entirely different tradition, the tradition of complex fluids. Here field equations are used to describe each of the force fields: stress stain mechanical relations, diffusion, electrical migration, and convection. Fields are combined by a variational approach like EnVarA [35-38] that guarantees mathematical consistency of the models chosen.

The key is to always make models of specific systems—including the apparatus and setup used to study them—and then to solve those models with systematic well defined approximations that other scientists and mathematicians can verify.

**Biology is easier than physics** in this particular case. In general, creating multiscale multifield models is a forbidding challenge, but fortunately one does not have to work in general if one is interested in engineering or biological systems.

Biology and engineering are rarely concerned with any possible system. They are mostly concerned with specific systems with specific structures that behave robustly when parameters are in certain limited ranges. These systems have a purpose and that requires them to follow macroscopic rules over a substantial range of conditions.

The design of the systems of biology and engineering can make analysis easier. In engineering, the purpose of systems is often obvious, and it is not necessary to consider systems in general. There is no need to study the operation of an automobile engine with water in the gas tank, or of an amplifier without a power supply.

In biology, evolution often goes down beaten paths, in which many complicated phenomena are restricted by structures, or by the limited range of conditions in which life exists.

Confining models to stay on these beaten paths focuses attention and makes possible what otherwise seems unapproachable. The Hodgkin Huxley treatment of the binary signal of nerve and muscle (now mysteriously called 'digital' although it does not involve fingers, or the numbers five or ten) is an example.[84] The hierarchy of models of the action potential reach from the atomic origin of its voltage sensor, through the channels that control current, to the current flow itself and how it produces a meter long signal. Biology allows analysis from atom size to arm length. A general analysis from Ångstroms to meters is made possible by structures at every scale. The enormous range and density of structures in biology creates a hierarchy in which analysis is possible.[85] Analysis that follows the path of those structures is following the path of natural selection. It, like the living beings it analyzes, can survive and succeed in an environment where general analysis is inconceivable.

**Setting boundaries**. The boundaries I propose for statistical mechanics are then easy to enumerate

1) Electrodynamics comes first because its equations are universal and exact when written in the form of the Core Maxwell Equations eq. (2)−(5).
2) Structures and boundaries must be involved, that describe the system and specific experimental setup used for measurement, albeit in an approximate way.
3) Systems with known function, of known structure, should be studied first. These often dramatically simplify problems, as they were designed to do, by engineers or evolution, once we known how to describe and exploit them with mathematics.
4) Systems that are devices, with well defined inputs, outputs, and input-output relations, should be identified because their properties are so much easier to deal with than systems and machines in general. Fortunately, devices are found throughout living systems, albeit not as universally (or as clearly defined) as in engineering systems.

When statistical mechanics is used without bounds, it is a quicksand which cannot support a hierarchy of models. Statistical mechanics without bounds is a dangerous foundation for structures with charge. They are likely to fail.

When statistical mechanics is used within bounds, the quicksand is constrained within walls, and the foundation and structures of our models can be strong and useful.

**Statistical mechanics within bounds can provide the foundation so badly needed for our models of biological and biochemical systems.**

Electrodynamics is always a safe foundation. Statistical mechanics can take its rightful place alongside electrodynamics once it is bound within structures and their boundary conditions.

### Acknowledgement.

Ardyth Eisenberg helped invent the title of this paper. She provided support and assistance that is beyond what can be described within the finite bounds of these words.